\documentclass[prl,amsmath,amssymb,twocolumn,showpacs,nofootinbib]{revtex4}
\usepackage{graphicx}
\usepackage{dcolumn}
\usepackage{bm}

\begin{document}

\title{Bloch-like Electronic Wave Functions in Two-Dimensional
  Quasicrystals} 

\author{Shahar Even-Dar Mandel}
\author{Ron Lifshitz}%
 \email{ronlif@tau.ac.il}
\affiliation{Raymond and Beverly Sackler School of Physics and
  Astronomy, Tel Aviv University, 69978 Tel Aviv, Israel} 

\date{\today}

\begin{abstract}
  Electrons in quasicrystals generically possess critical wave
  functions that are neither exponentially-localized nor extended, but
  rather decay algebraically in space. Nevertheless, motivated by
  recent calculations on the square and cubic Fibonacci quasicrystals,
  we investigate whether it is possible to obtain extended wave
  functions expressed as linear combinations of degenerate, or
  nearly-degenerate, critical eigenfunctions. We find that not only is
  this possible, but that the wave functions that emerge are
  Bloch-like, exhibiting the quasiperiodic long-range order of the
  underlying quasicrystal. We discuss the significance of this result
  for the study of electronic properties of real quasicrystals.
\end{abstract}


\maketitle

Although research in quasicrystals celebrated its twenty-fifth
anniversary last year~\cite{shechtman,jubilee}, many gaps still remain
in the fundamental understanding of their electronic
properties~\cite{mayou08}. A major shortcoming is the lack of a
quasiperiodic analog of the Bloch theorem, which plays an important
role in the study of the electronic properties of periodic crystals,
serving as the basis for band-structure and electronic-transport
theories. Naively, one would expect to encounter a quasiperiodic
version of the Bloch theorem when dealing with
quasicrystals~\cite{sinai,birman}, predicting the existence of
extended eigenfunctions that display the underlying quasiperiodic
long-range order. Some experiments even seem to indicate that such
extended eigenfunctions exist in real quasicrystals~\cite{rotenberg}.
However, as previous studies have
shown~\cite{review1,review2,review3}, the electronic eigenfunctions in
quasicrystals are generically found to be critical, with algebraic
spatial decay.
 
We have recently reintroduced the use of separable models---the square
and cubic Fibonacci tilings~\cite{squarefib}---for studying the
physical properties of quasicrystals (for precise definitions of the
terms `crystal' and `quasicrystal' see~\cite{definition1}
and~\cite{definition2}). These models allow us to use well-known
results for the 1-dimensional ($1d$) Fibonacci quasicrystal and extend
them to higher dimensions. Our past studies of the electronic energy
spectra of these quasicrystals~\cite{ilan,me,me2008} led us to
conjecture (as we explain below) that linear combinations of
degenerate eigenfunctions, all of which are critical, may produce
extended wave-functions, but we failed to find such combinations. In
this Letter, using a new iterative search algorithm formulated by
Elser \emph{et al.}~\cite{elser}, we establish that not only do such
wave-functions exist, they actually exhibit Bloch-like properties.

\begin{figure}
\begin{center}
\scalebox{0.3}{\includegraphics{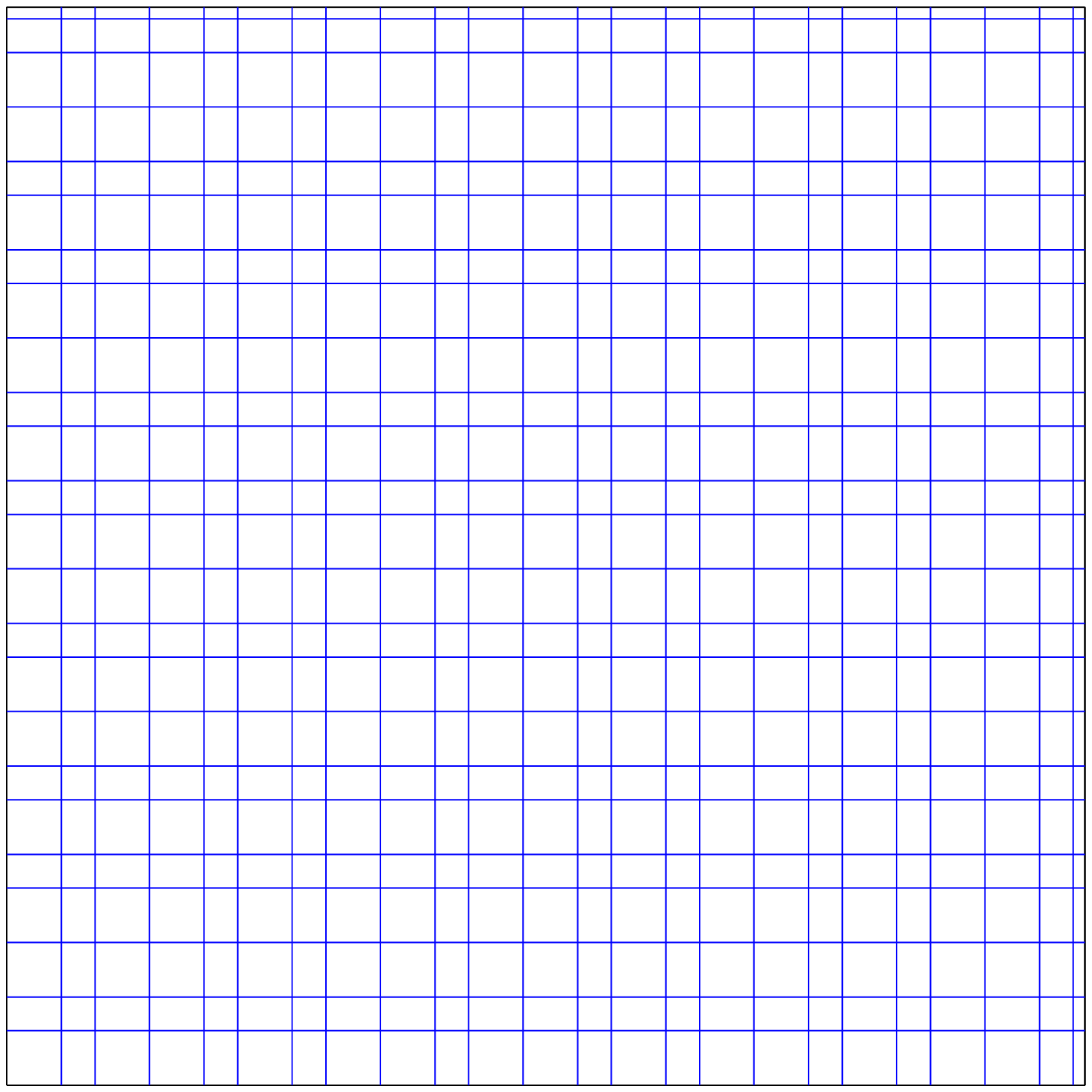}}
\caption{\label{fig:patch} A section of the square Fibonacci tiling,
  whose geometric properties are described in Ref.~\cite{squarefib}.}
\end{center}
\end{figure}

The square Fibonacci tiling is constructed by superimposing two $1d$
Fibonacci grids perpendicular to each other, as shown in
Fig.~\ref{fig:patch}. The off-diagonal tight-binding model on this
quasicrystal~\cite{ilan,me,me2008} assumes equal on-site energies
(taken to be zero), and hopping that is restricted along tile edges,
with amplitude 1 for long edges and $T$ for short edges, where we take
$T\geq 1$. The Schr\"odinger equation is then given by
\begin{eqnarray}
  \label{eq:twoDeq}\nonumber
  &&T_{i+1} \Psi(i+1,j) + T_i \Psi(i-1,j)\\
  &&+ T_{j+1} \Psi(i,j+1) + T_j \Psi(i,j-1)
  = \epsilon\Psi(i,j),
\end{eqnarray}
where $\Psi(i,j)$ is the value of a $2d$ eigenfunction on a vertex
labeled by the two integers $i$ and $j$, and $\epsilon$ is the
corresponding eigenvalue. The hopping amplitudes $T_j$ are equal to 1
or $T$ according to the Fibonacci sequence
$\{T_j\}=\{1,T,1,1,T,1,T,1,1,T,1,1,T,1,T,\ldots\}$. The fact that
Eq.~(\ref{eq:twoDeq}) is separable allows one to use the known
solutions for the $1d$
problem~\cite{1dmodel1,1dmodel2,1dmodel3,1dmodel4,review1,review2,%
  review3,review4} in order to study higher-dimensional models (as was
done for similar models in the
past~\cite{2dmodel1,2dmodel2,2dmodel3,2dmodel4,2dmodel5,2dmodel6,ashraff}).
Two-dimensional eigenfunctions are therefore Cartesian products of
$1d$ eigenfunctions, $\Psi_{n,m}=\psi_n\bigotimes\psi_m$, and the
corresponding $2d$ eigenvalues are the pairwise sums of $1d$
eigenvalues $\epsilon_{n,m}=\epsilon_n+\epsilon_m$. The $1d$
eigenvalue problem is solved by considering periodic approximants
of the $1d$ Fibonacci quasicrystal. The unit cell of the $N^{th}$
order approximant consists of $F_N$ sites---where $F_N$ is the
$N^{th}$ Fibonacci number, defined by $F_N=F_{N-1}+F_{N-2}$ with
$F_0=F_1=1$---and therefore its spectrum consists of $F_N$ bands.
Explicit diagonalization of the Hamiltonian for a single $1d$ unit
cell, with periodic or antiperiodic boundary conditions, yields the
edges of these bands, and hence provides the full structure of the
$1d$ spectrum.

Our studies of the energy spectrum of the $2d$ model revealed that for
$T$ close to 1 ({\it i.e.} near the periodic limit) the addition of
$1d$ spectra gives rise to a $2d$ spectrum containing energy
intervals, similar in its properties to the spectrum of a periodic
crystal, even though the $1d$ spectrum is nowhere dense, containing no
interval. This behavior of the spectrum led us to expect that
eigenfunctions would be extended, even though they are constructed as
Cartesian products of critical eigenfunctions. We proposed that the
degeneracy of the $2d$ spectrum, at values of $T$ close to 1, may
provide a mechanism for the emergence of extended wave functions as
the dimensionality of the quasicrystal increases. Our conjecture was
based on the idea that appropriate linear combinations of degenerate
algebraically-decaying eigenfunctions, centered about different sites
but with sufficient overlap, may yield an extended function that spans
the whole quasicrystal. To test this hypothesis we used an \emph{ad
  hoc} approach to try to find an extended linear combination of
zero-energy eigenfunctions~\cite{me}. The fact that the zero-energy
eigenfunctions are macroscopically-degenerate, gave us great freedom,
yet we only managed to obtain wave functions that were extended along
one direction, but very strongly localized in the perpendicular
direction.\footnote{Note that in Ref.~\cite{ilan}---where we showed
  that for every $1d$ eigenfunction $\psi_n(i)$ there exists a $2d$
  eigenfunction $\Psi_n(i,j)=(-1)^j\psi_n(i)\psi_n(j)$ with zero
  energy---we gave an incorrect estimate of the spatial extent of
  $\Psi_n$.}

The iterative search algorithm of Elser \emph{et al.}~\cite{elser}
gives us a new way of maximizing the extent of wave functions that are
spanned by a given set of eigenfunctions. The algorithm is based on
formulating a problem as an attempt to simultaneously satisfy two
constraints, representing them geometrically, and using projections
iteratively between the representations until convergence is obtained.
In our problem, the first constraint is for the function to be spanned
by the given set $\{\Psi_{n.m}\}$ of degenerate eigenfunctions, which
determines a vector subspace and the natural projection operator into
it. The second constraint is for the wave function to be totally
extended, with equal amplitudes on all lattice sites. This constraint
determines a projection operator in which the amplitudes of a wave
function are all set to be equal, while its phases are kept unchanged.
The two operators used for the application of the algorithm are thus
\begin{eqnarray}
P_1(\Psi)&=&\sum_{n,m}\langle\Psi_{n.m}|\Psi\rangle\Psi_{n,m},\\
P_2(\Psi)(j,k)&=&\frac{\Psi(j,k)}{F_N|\Psi(j,k)|},
\end{eqnarray}
and the search is stopped after applying the projection operator
$P_1$, ensuring that the final wave function can be spanned by the
required set of eigenfunctions, although generally it will not be
totally-extended. 

\begin{figure}
\begin{center}
\scalebox{0.4}{\includegraphics{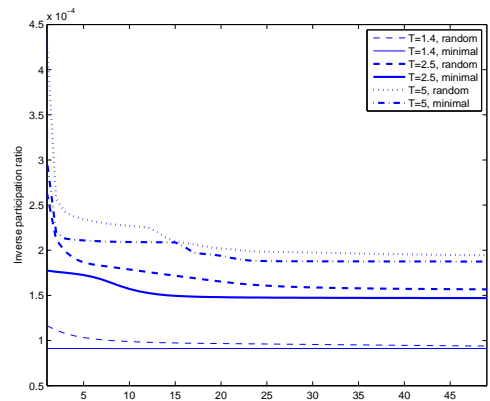}}
\caption{\label{fig:convergence} The inverse participation ratio as a
  function of iteration steps for different values of the relative
  hopping amplitude $T$ and different types of initial guesses.} 
\end{center}
\end{figure}

The algorithm minimizes the $\ell_2$ distance from the set of
totally-extended wave functions. However, a more common measure for
the extent of a wave function is its inverse participation ratio,
\begin{equation}\label{eq:ipr_def}
I\{\Psi\} =
\frac{\sum_{j,k}|\Psi(j,k)|^4}{\left(\sum_{j,k}|\Psi(j,k)|^2\right)^2}, 
\end{equation}
which we use here to trace the convergence of the algorithm.  We find
that the iterative search algorithm quite easily succeeds in finding
extended linear combinations for all three kinds of degenerate sets
considered below. Fig.~\ref{fig:convergence} shows the inverse
participation ratio as a function of iteration number for some typical
examples. The algorithm is seen rapidly to converge to a steady,
minimal, value of the inverse participation ratio, corresponding to a
maximally-extended wave function. Naively, one may expect the
convergence to accelerate if one picks the most extended eigenfunction
as an initial guess, but it turns out that for low values of $T$ the
algorithm cannot escape the neighborhood of such initial guesses. In
such cases, an extended function with random phases, taken as the
initial guess, yields a better final outcome, as demonstrated in
Fig.~\ref{fig:convergence}. 

Turning to actual results, we start by reexamining the possibility of
generating an extended wave function by using the
macroscopically-degenerate zero-energy eigenvalue.  Applying the
iterative search algorithm to the set of zero-energy eigenfunctions
yields highly-extended eigenfunctions, despite our past
failure~\cite{me}, as can be seen in the supplementary
material~\cite{epaps}.

The high degeneracy of zero-energy eigenfunctions is an artifact of
our model which does not have any real physical significance. We
therefore consider non-zero energies $\epsilon\geq 1+T$, taken from
the upper quarter of the $2d$ energy spectrum, which is
fully-contained within the interval $-2-2T$ to $2+2T$. It is possible
for certain $2d$ eigenvalues to be obtained as sums of pairs of $1d$
eigenvalues in a number of different ways, and thus have nontrivial
degeneracy. To approximate this degeneracy for finite order
approximants, we consider for each energy in the $2d$ spectrum the
number of bands---constructed as sums of pairs of $1d$ bands---in
which it is contained. We then pick a representative eigenfunction
from each of these bands to be included in the degenerate set.
Applying the algorithm on such sets of degenerate eigenfunctions
yields extended wave functions with some appearance of long-range
order, as can be seen in the supplementary material~\cite{epaps}.

\begin{figure}
\begin{center}
\scalebox{0.7}{\includegraphics{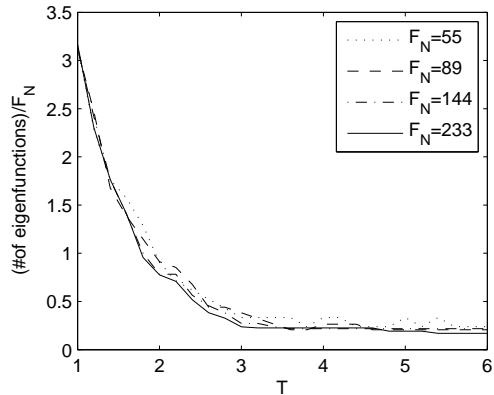}}
\caption{\label{fig:degn} The largest number of bands with a common
  energy value divided by $F_N$ as a function of the relative hopping
  parameter $T$, for different orders of approximants. The convergence
  of the curves implies that the spatial density of degenerate
  eigenfunctions, per unit area, decreases as the order of approximant
  increases.}
\end{center}
\end{figure}

\begin{figure*}[bt]
\begin{center}
 \scalebox{0.4}{\rotatebox{00}{\includegraphics*{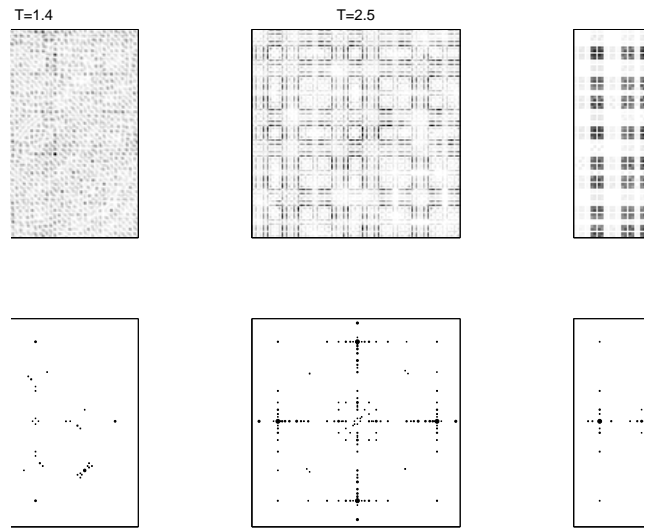}}}
 \caption{\label{fig:blochs} The intensities $|\Psi|^2$ of the
   wave-functions obtained by the search algorithm on a $144\times
   144$ approximant for different values of $T$ are shown in the upper
   row. The underlying Fibonacci order becomes evident as $T$
   increases. This is supported by the emergence of Bragg peaks in the
   diffraction patterns, shown on the bottom row. Bragg peaks are
   represented by circles whose radius is proportional to their
   amplitude. Additional images and animations can be found in the
   supplementary material~\cite{epaps}.}
\end{center}
\end{figure*}

Despite this apparent success, one should be careful in assigning it
any physical significance, because for a real electron to be able to
explore the degenerate set of eigenfunctions these functions must
possess sufficient spatial overlap. Fig.~\ref{fig:degn} plots the
number of bands that contain the most degenerate energy, $\epsilon\geq
1+T$, as a function of $T$, divided by the corresponding Fibonacci
number $F_N$ for four consecutive orders of approximants. The steep
decline of the curves as $T$ increases shows that degeneracy is
negligible for high values of $T$, as expected. The convergence of the
curves for increasing $N$ implies that the degeneracy is proportional
to $F_N$ whereas the number of sites is equal to $F_N^2$. Thus, as the
order of approximant increases, the spatial density of degenerate
eigenfunctions decreases.  In the quasiperiodic limit the degenerate
eigenfunctions might only have negligible overlap, and a real electron
may retain the critical nature of its eigenfunction.

Finally, we consider the most physically-relevant situation in which
thermal fluctuations, a small amount of disorder, or the existence of
inhomogeneous external fields, impose a small uncertainty in the exact
energy of an electron. We take this uncertainty to be
$\Delta\epsilon=0.01$, which is less than a quarter of one percent of
the total bandwidth. We treat all the eigenfunctions with eigenvalues
within a window of size $\Delta\epsilon$ as if they were degenerate,
and apply the iterative search algorithm to the largest
nearly-degenerate set, with energies above $1+T$. The upper row of
images in Fig.~\ref{fig:blochs} show the intensities of the wave
functions that are obtained by the algorithm, for three values of $T$.
Even though the algorithm is devised only to maximize the extent of
the wave functions, regardless of their actual structure, for
sufficiently strong quasiperiodicity, or large $T$, it gives rise to
the emergence of Bloch-like wave functions.

For relatively low values of $T$ the spatial decay of the critical
eigenfunctions is relatively slow and the overlap between them is such
that the wave functions obtained by the iterative algorithm appear to
be smeared over the entire approximant with very weak apparent
long-range order. As $T$ increases the spatial decay of the critical
eigenfunctions becomes more dominant, and the obtained wave functions
display Bloch-like behavior, as can be seen in the plots for $T=2.5$
and $T=5$. These wave functions are Bloch-like in the sense that their
intensities $|\Psi|^2$ exhibit the underlying quasiperiodic order of
the quasicrystal, although one cannot associate with them a
well-defined non-reciprocal-lattice wave-vector, which is an
additional requirement of the Bloch theorem (see
Refs.~\cite{lifshitzsymmetry} and \cite{lifshitzbravais} for a
discussion of reciprocal lattices of quasicrystals).

This observation is supported by the Fourier transforms of the
probability densities $|\Psi|^2$, shown at the bottom of
Fig.~\ref{fig:blochs}. Each Bragg peak with amplitude above some
threshold is represented by a circle whose radius is proportional to
the amplitude.  While some Bragg peaks appear already for $T=1.4$ they
lack square symmetry, and the diffraction contains a considerable
diffuse component (not shown due to the threshold).  For higher values
of $T$ it clearly shows Bragg peaks in a pattern similar to that of
the square Fibonacci quasicrystal itself.

\begin{figure}
\begin{center}
\scalebox{0.5}{\rotatebox{00}{\includegraphics*{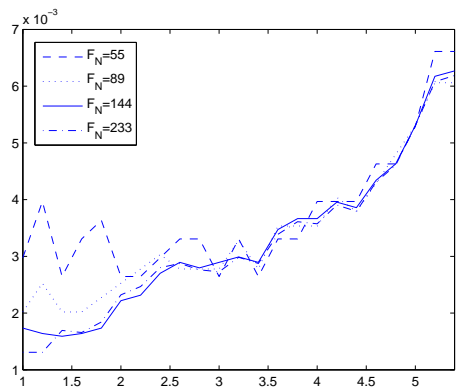}}}
\caption{\label{fig:degnn} The number of eigenfunctions in the most
  populated energy interval, in the upper half of the $2d$ spectrum,
  divided by the number of sites in the $2d$ quasicrystal ($F_N^2$) as
  a function of $T$. The convergence of the curves implies that the
  spatial density of nearly-degenerate eigenfunctions does not change
  as the order of approximant increases.}
\end{center}
\end{figure}

To see whether these results are valid in the quasiperiodic limit
Fig.~\ref{fig:degnn} plots the number of eigenfunctions in the most
populated energy interval $\Delta\epsilon$ as a function of $T$ for
approximants of order $9-12$ ($F_N=55-233$). The number of degenerate
eigenfunctions is divided by the number of sites, $F_N^2$, and can be
seen to scale with it by the convergence of the different curves. This
implies that the spatial density of nearly-degenerate eigenfunctions
remains constant, so that appropriate linear combinations can retain
the Bloch-like nature as one approaches the quasiperiodic limit.

The existence of Bloch-like wave functions in the square Fibonacci
quasicrystal encourages us to seek out similar results in more
realistic models of quasicrystals, like the Penrose tiling. Even
though individual eigenfunctions are critical in these models, we
intend to investigate whether combinations of nearly-degenerate
eigenfunctions can be extended. These nearly-degenerate combinations
of functions may turn out to play an important role in realistic
situations, significantly affecting the transport properties and
quantum dynamics of electronic wave packets in real quasicrystals.

This research is supported by the Israel Science Foundation
through Grant No.~684/06.

\bibliography{bloch}

\end{document}